\documentclass{article}

\usepackage[utf8]{inputenc}
\usepackage{microtype}
\usepackage{graphicx}
\usepackage{subfigure}
\usepackage{booktabs} 
\usepackage{multirow}

\usepackage{hyperref}

\usepackage[accepted]{icml2018}

\icmltitlerunning{Attention as a Perspective for Learning Tempo-invariant Audio Queries}

\begin{document}

\twocolumn[
\icmltitle{Attention as a Perspective for Learning Tempo-invariant Audio Queries}




\begin{icmlauthorlist}
\icmlauthor{Matthias Dorfer}{aff1}
\icmlauthor{Jan Haji{\v{c}} jr.}{aff2}
\icmlauthor{Gerhard Widmer}{aff1}
\end{icmlauthorlist}

\icmlaffiliation{aff1}{Johannes Kepler University Linz, Austria}
\icmlaffiliation{aff2}{Charles University, Czech Republic}

\icmlcorrespondingauthor{Matthias Dorfer}{matthias.dorfer@jku.at}

\icmlkeywords{Machine Learning, ICML}

\vskip 0.3in
]



\printAffiliationsAndNotice{}  

\begin{abstract}
Current models for audio--sheet music retrieval
via multimodal embedding space learning 
use convolutional neural networks
with a fixed-size window for the input audio.
Depending on the tempo of a query performance,
this window captures more or less musical content,
while notehead density in the score is largely
tempo-independent.
In this work we address this disparity 
with a soft attention mechanism,
which allows the model
to encode only those parts of an audio excerpt
that are most relevant 
with respect to efficient query codes.
Empirical results on classical piano music
indicate that attention is beneficial 
for retrieval performance, 
and exhibits intuitively appealing behavior.
\end{abstract}

%
\section{Introduction}
\label{sec:introduction}
Cross-modal embedding models have demonstrated
the ability to retrieve sheet music using an audio query, and vice
versa, based on just the raw audio and visual
signal \citep{Dorfer2017Audio2Score}. 
A limitation of the system was that
the field of view into both modalities had a fixed size.
This is most pronounced for audio: a human listener can easily recognize
the same piece of music played in very different tempi,
but when the audio is segmented into spectrogram excerpts
with a fixed number of time steps, these contain disparate
amounts of musical content, relative to what the model
has seen during training. The tempo can also
change within a single query, especially in live retrieval
settings. 

We propose applying an attention mechanism
\citep{olah2016attention,DBLP:conf/icassp/ChanJLV16,DBLP:journals/corr/BahdanauCB14,DBLP:conf/nips/MnihHGK14,DBLP:conf/icml/XuBKCCSZB15,NIPS2017_7181,DBLP:conf/ismir/SouthallSH17}
over the audio input, to distinguish parts
of the audio that are in fact useful 
for finding the corresponding sheet music snippets.
The system 
can
then
adapt to tempo changes: both to lower
and to higher densities of musical events.
Our experiments show that attention
is indeed a promising way to obtain tempo-invariant
embeddings for cross-modal retrieval.

%
\section{Audio--Sheet Music Embedding Space Learning with Attention}
\label{sec:methods}
%

We approach the cross-modal audio-sheet music retrieval problem
by learning a low-dimensional multimodal embedding space (32 dimensions)
for both snippets of sheet music and excerpts of music audio.
We desire for each modality a projection into
a shared space where
semantically similar items of the two modalities 
are projected close together, and
dissimilar items far apart.
Once the input modalities are embedded in such a space, 
cross-modal retrieval is performed
using simple distance measures and nearest-neighbor search.
\begin{figure}[t!]
 \centerline{\includegraphics[width=0.94\columnwidth]{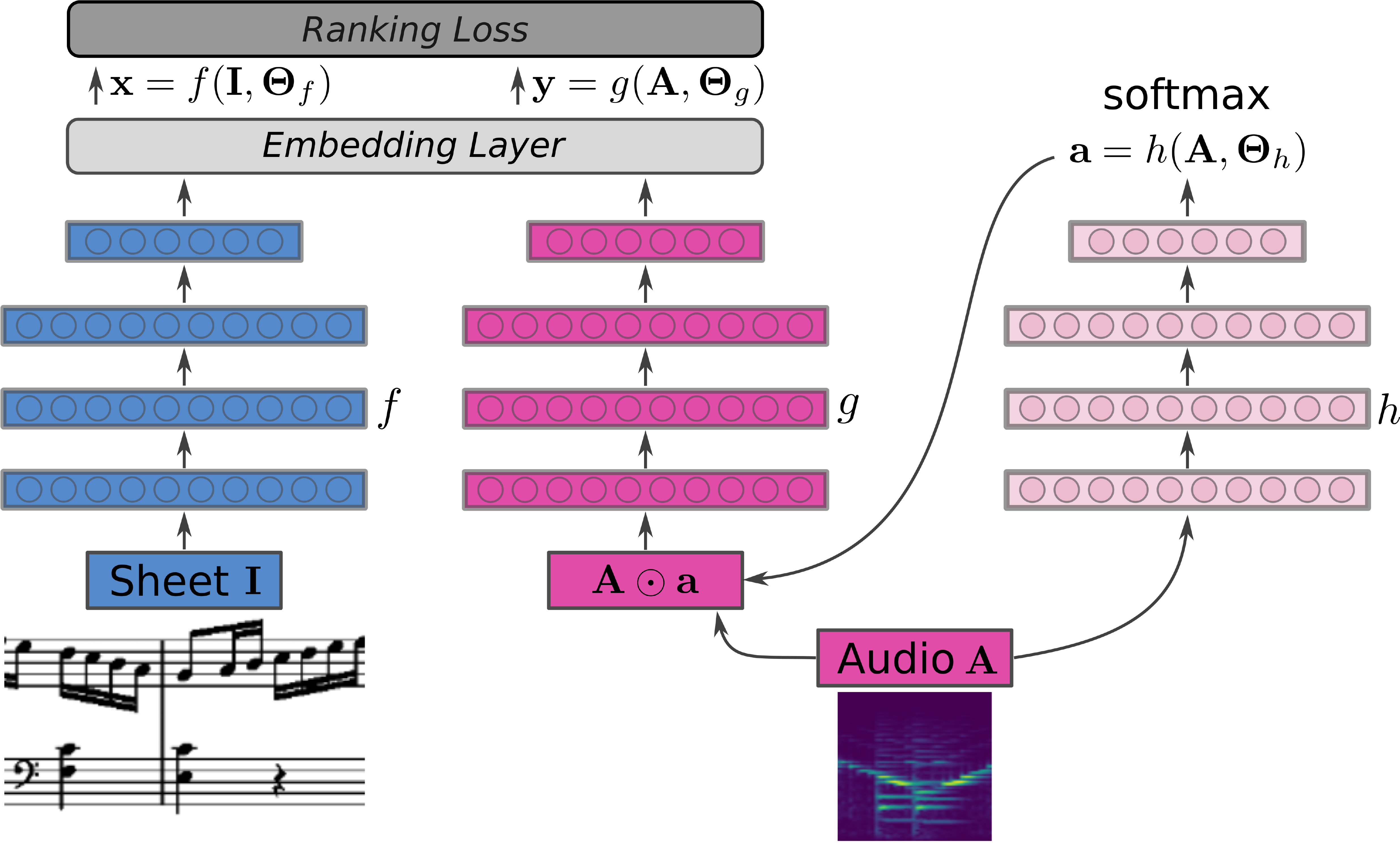}}
 \caption{Audio--sheet music embedding space learning with soft-input-attention on the input audio.}
\label{fig:model}
\end{figure}
\begin{figure*}[t!]
 \centerline{\includegraphics[width=2.0\columnwidth]{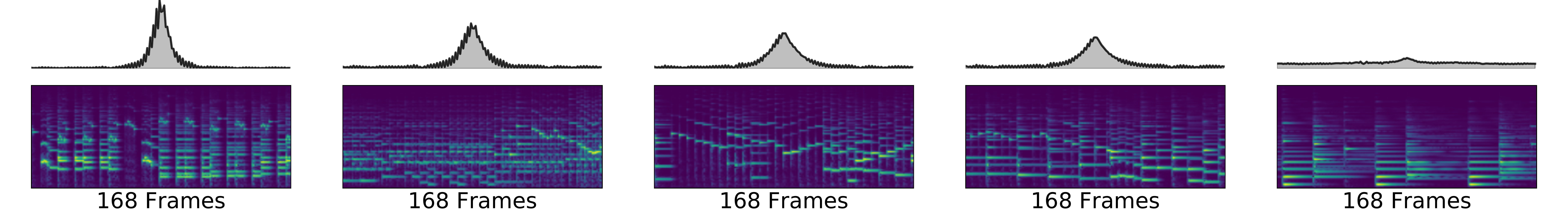}}
 \caption{Audio queries (all 168 frames) and corresponding attention vectors of model \emph{BL + AT + LC}.
 Note, that when the music gets slower (see right most example) and covers less onsets,
 the attention mechanisms starts to consider a larger temporal context.}
\label{fig:attention_samples}
\end{figure*}

We train the embedding space using convolutional neural networks;
Figure \ref{fig:model} sketches the network architecture.
The baseline model (without attention)
consists of two convolutional pathways:
one is responsible for embedding the sheet music,
and the other for embedding the audio excerpt.
The key part of the network is the canonically correlated embedding layer \citep{Dorfer2018CCALayer},
which forces the two pathways to learn representations that
can then be projected into a shared space 
(and finds these projections);
the desired properties of this multimodal embedding space
are enforced by training with pairwise ranking loss \citep{kiros2014unifying}.
This is the basic structure of the model recently described and evaluated in \cite{Dorfer2017Audio2Score}.
This attention-less model
serves as a baseline in our experiments.

As already mentioned, this model trains and operates on fixed-size
input windows from both modalities. At runtime, the input audio (or sheet music)
query therefore has to be broken down into excerpts of the given size.
When processing audio played in different tempi,
the fixed-size excerpts contain significantly less (or more) 
musical content -- esp. onsets -- than excerpts that the model has been trained
on. One may of course combat this with data augmentation, but a more
general solution is to simply let the model read as much information from
the input excerpt as it needs. 

We explore using a {\bf soft attention mechanism} for this purpose.
First, we substantially increase the audio field of view 
(number of spectrogram frames), up to a factor of four.
Next, we add to the model the attention pathway $h$,
which is implemented as a softmax layer that outputs a weight $a_t$ for each
input spectrogram frame in $\mathbf{A}$.
Before feeding the spectrogram to the audio embedding network $g$,
we multiply each frame with its attention weight.
This enables the model to cancel out irrelevant parts of the query.

%
\section{Experimental Evaluation and Discussion}
\label{sec:experiments}
In our retrieval experiments, we use a dataset of classical piano music, \emph{MSMD} \citep{Dorfer2018Audio2Score}.
It contains 479 pieces of 53 composers,
including Bach, Mozart, Beethoven and Chopin,
totalling 1,129 pages of sheet music
and 15+ hours of audio, with fine-grained 
cross-modal alignment between note onsets and noteheads.
The scores and audio are both synthesized
based on LilyPond, but results 
in \citep{Dorfer2018Audio2Score} suggest
that the embedding models trained on this data do
generalize to real scores and performances.
Our experiments are carried out on 
aligned
snippets of sheet music and spectrogram excerpts,
as indicated in Figure~\ref{fig:model}.
Given an audio excerpt as a search query,
we aim to retrieve (only) the corresponding snippet 
of sheet music of the respective piece.

As an experimental baseline, we use a model similar to the one presented in \citep{Dorfer2017Audio2Score}, which does not
use attention (denoted as \emph{BL}).
The second model we consider follows exactly the same architecture,
but is additionally equipped with the soft attention mechanism
described above (\emph{BL + AT}).
The temporal context for both models is 84 frames ($\approx$ 4 seconds), twice as much compared to \cite{Dorfer2018Audio2Score}.
The third model is the same as \emph{BL + AT}
but is given a larger temporal context (168 frames) for the audio spectrogram (\emph{BL + AT + LC}).
This should reveal if and how 
the model will learn to focus on
the relevant parts in the audio,
depending on the musical content it is presented with.
The sheet music snippet has the same dimensions ($80 \times 100$ pixels) 
for all models,
implying that the audio network has to adapt to this fixed condition.
As evaluation measures we compute the \emph{Recall@k (R@k)},
the \emph{Mean Reciprocal Rank (MRR)}, as well as the \emph{Median Rank (MR, low is better)}.

\begin{table}[t!]
 \footnotesize
 \begin{center}
 \begin{tabular}{rccccc}
 \toprule
 \textbf{Model} & \textbf{R@1}  & \textbf{R@5} & \textbf{R@25} & \textbf{MRR} & \textbf{MR} \\
 \midrule
BL  			& 41.4 & 63.8 & 77.2 & 51.8 & 2 \\	
BL + AT  		& 47.6 & 68.2 & 79.4 & 57.1 & 2 \\	
BL + AT + LC	& 55.5 & 77.1 & 85.8 & 65.1 & 1 \\	
 \bottomrule
 \end{tabular}
\end{center}
 \caption{Comparison of retrieval results (10,000 candidates).}
 \label{tab:eval_methods}
\end{table}
Table \ref{tab:eval_methods} summarizes our results.
The attention mechanism (\emph{BL + AT}) improves the retrieval performance over the baseline consistently across all retrieval metrics.
With increased temporal context of the attention model 
(\emph{BL + AT + LC})
we achieve another substantial jump in performance.

To investigate whether the attention mechanism behaves 
according to our intuition, 
we plot audio queries along with their attention weights in Figure~\ref{fig:attention_samples}.
Depending on the spectrogram content, the model indeed attends
to whatever it believes is a representative counterpart of the target sheet music snippet.
Since the fixed-size sheet snippets contain roughly similar 
amounts of notes, as the density of noteheads
is independent on the tempo of the piece, attention is
sharply peaked
when the density of notes in the audio is high,
and conversely it is distributed more evenly
when there are fewer notes in the audio excerpt.

Given the improved retrieval performance and the intuitive behavior
of the model, we think this is a promising line of research
for reducing the sensitivity of cross-modal music retrieval models
to the audio input window size.
By extension, this is a step towards tempo-invariant representations,
at least in the context of retrieval.

\section*{Acknowledgements}
This research was supported in part by the European Research
Council (ERC) under grant ERC-2014-AdG 670035
(ERC Advanced Grant, project "Con Espressione").
Jan Haji{\v{c}} jr. wishes to acknowledge support by the Czech
Science Foundation grant no. P103/12/G084 and Charles University
Grant Agency grant no. 1444217.

\bibliography{bibliography}
\bibliographystyle{icml2018}

\end{document}